\documentclass[twocolumn]{aastex62}

\usepackage{amsmath}

\graphicspath{{./}{figures/}}

\shorttitle{the BNS merger / AIC FRB Channel}
\shortauthors{Margalit, Berger \& Metzger}

\begin{document}

\title{Fast Radio Bursts from Magnetars Born in Binary Neutron Star Mergers and Accretion Induced Collapse}

\email{benmargalit@berkeley.edu}

\author{Ben Margalit}
\altaffiliation{NASA Einstein Fellow}
\affil{Astronomy Department and Theoretical Astrophysics Center, University of California, Berkeley, Berkeley, CA 94720, USA}

\author{Edo Berger}
\affil{Center for Astrophysics, Harvard \& Smithsonian, 60 Garden Street, Cambridge, MA 02138, USA}

\author{Brian D. Metzger}
\affil{Department of Physics and Columbia Astrophysics Laboratory, Columbia University, 538 West 120th St., New York, NY 10027}


\begin{abstract}

Recently born magnetars are promising candidates for the engines powering fast radio bursts (FRBs).  The focus thus far has been placed on millisecond magnetars born in rare core-collapse explosions, motivated by the star forming dwarf host galaxy of the repeating FRB\,121102, which is remarkably similar to the hosts of superluminous supernovae (SLSNe) and long gamma-ray bursts (LGRB).  However, long-lived magnetars may also be created in binary neutron star (BNS) mergers, in the small subset of cases with a sufficiently low total mass for the remnant to avoid collapse to a black hole, or in the accretion-induced collapse (AIC) of a white dwarf.  A BNS FRB channel will be characterized by distinct host galaxy and spatial offset distributions than the SLSNe/LGRB channel; we anticipate a similar host population, although possibly different offset distribution for AIC events.  We show that both the BNS and AIC channels are consistent with the recently reported FRB~180924, localized by ASKAP to a massive quiescent host galaxy with an offset of about 1.4 effective radii.  FRBs from magnetars born in BNS mergers and AIC will be accompanied by persistent synchrotron radio emission on timescales of months to years, powered by the nebula of relativistic electrons and magnetic fields inflated by the magnetar flares, or on longer timescales through interaction of the merger ejecta with the interstellar medium; this timescale is shorter than for the SLSN/LGRB channel.  Using models calibrated to FRB\,121102, we make predictions for the dispersion measure, rotation measure, and persistent radio emission from magnetar FRB sources born in BNS mergers or AIC, and show these are consistent with upper limits from FRB\,180924 for reasonable parameters.  Depending on the rate of AIC, and the fraction of BNS mergers leaving long-lived stable magnetars, the birth rate of repeating FRB sources associated with older stellar populations could be comparable to that of the core-collapse channel.  We also discuss potential differences in the repetition properties of these channels, as a result of differences in the characteristic masses and magnetic fields of the magnetars.

\end{abstract}


\section{Introduction} 
\label{sec:intro}

Fast radio bursts (FRB) are millisecond duration pulses of coherent radio emission with large dispersion measures (DM) well above the contribution from the Milky Way, thus implicating an extragalactic origin (e.g., \citealt{Lorimer+07,Keane+12,Thornton+13, Spitler+14,Ravi+15,Petroff+16,Champion+16,Lawrence+17,Shannon+18}; see \citealt{PetroffHessels&Lorimer19,Cordes&Chatterjee19} for recent reviews).  A cosmological origin was directly confirmed for the repeating FRB~121102 \citep{Spitler+14,Spitler+16} thanks to its precise localization to a low metallicity dwarf star forming galaxy at a redshift of $z=0.193$ \citep{Chatterjee+17,Tendulkar+17}.  FRB~121102 is also spatially coincident with a compact ($<0.7$ pc) and luminous ($\nu L_{\nu}\sim 10^{39}$\,erg\,s$^{-1}$) persistent radio synchrotron source \citep{mph+.2017}.  It exhibits an enormous rotation measure, ${\rm RM}\sim 10^{5}$ rad m$^{-2}$ (\citealt{Michilli+18}; see also \citealt{Masui+15}), which exceeds those of other known astrophysical sources, with the exception of Sgr~A* and the flaring magnetar SGR J1745-2900 located in the Galactic Center \citep{Eatough+13}.   

Recently, the Australian Square Kilometre Array Pathfinder (ASKAP) localized a second event --- FRB~180924 --- based on the detection of a single burst \citep{Bannister+19}.  This FRB has not exhibited repetition so far, it is located in a more massive and quiescent galaxy than that of FRB~121102, it is not accompanied by a persistent radio source to a limit of $\sim3$ times lower than FRB~121102, and it exhibits a low RM of $14$~rad~m$^{-2}$ \citep{Bannister+19}.

Although dozens of models have been proposed for FRBs, most are ruled out by a repeating, cosmological source like FRB~121102, while others are ruled out by the large all-sky rate {\it if the bulk of FRBs are non-repeating}. One compelling model for repeating FRBs are bursts generated by a young flaring magnetar \citep{Popov&Postnov13,Lyubarsky14,Kulkarni+14,Katz16,Lu&Kumar16,Metzger+17,Nicholl+17c,Kumar+17,Beloborodov17,Lu&Kumar18}.  This idea is supported by the atypical properties of the host galaxy of FRB~121102, which are similar to those of superluminous supernovae (SLSNe) and long gamma-ray bursts (LGRBs; \citealt{Tendulkar+17,Metzger+17,Nicholl+17c}), rare explosions that are powered by engines \citep{MacFadyen&Woosley99,Kasen&Bildsten10,Nicholl+17}. In such a model, the persistent radio source associated with FRB~121102 could be understood as emission from a compact magnetized nebula surrounding the young (decades to centuries old) magnetar and embedded in the expanding supernova (SN) ejecta \citep{Metzger+17,Kashiyama&Murase17,Omand+18,Margalit&Metzger18}.  The nebula is powered by nearly continual energy release from the magnetar, likely during the same sporadic flaring events responsible for the repeated FRBs \citep{Beloborodov17,Metzger+19}.  As shown by \citet{Margalit&Metzger18}, the radio flux of the nebula, and its large but decreasing RM, can both be explained in this model.  This notion is also supported by the recent detection of a radio source coincident with the SLSN PTF10hgi about 7.5 years post-explosion \citep{Eftekhari+19}.

However, SLSNe/LGRBs are not the only potential sites for magnetar birth.  The gravitational-wave (GW) driven merger of binary neutron stars (BNS) is generally believed to give rise to a massive magnetized neutron star (NS) remnant (e.g., \citealt{Price&Rosswog06,Bucciantini+12,Giacomazzo&Perna13}).  In most cases this remnant is well above the Tolman-Oppenheimer-Volkoff (TOV) maximum stable mass, $M_{\rm TOV}$, and thus is only temporarily stabilized against gravitational collapse by rapid rotation; once this rotational support is removed, by a combination of internal (``viscosity'') and external (e.g., magnetic dipole losses) stresses, the magnetar collapses to a black hole.  However, depending on the cosmic population of binary NSs, the uncertain value of $M_{\rm TOV}$, and the amount of mass loss during the merger, a modest fraction of mergers could lead to indefinitely stable NSs (e.g., \citealt{Metzger+08,Piro+17,Margalit&Metzger19}).  Such a stable magnetar may be similar to those formed in SLSNe and/or LGRBs, but will be surrounded by a much smaller and faster-expanding ejecta shell, and will be hosted by a different galaxy population.  These magnetars may also exhibit large spatial offsets from their host galaxies, as observed for short gamma-ray bursts (SGRBs; e.g., \citealt{Berger14}). If such magnetars give rise to FRBs, they will therefore be accompanied by different nebulae and host galaxy demographics (e.g., \citealt{Nicholl+17c,Yamasaki+18}). 

Another potential formation channel for magnetars is the accretion-induced collapse (AIC) of a white dwarf.  This can occur either due to accretion from a non-degenerate binary companion (e.g., \citealt{Nomoto+79,Taam&vandenHuevel86,Canal+90,Nomoto&Kondo91,Tauris+13,Schwab+15,Brooks+17}) or following the merger of two white dwarfs in a binary system (e.g., \citealt{Yoon+07,Schwab+16}).  As with BNS mergers, AIC is expected to occur in a range of host galaxy types due to the delay after star formation. However, due to the small natal kicks of the NSs formed through this channel, they may not occur with as large offsets as magnetars formed via BNS mergers.
AIC has been previously suggested as a possible progenitor for FRB~121102 \citep{Kashiyama&Murase17,Waxman17}, however these models differ significantly from the flaring magnetar model discussed here.

Here, we develop predictions of the BNS merger and AIC magnetar channel for FRBs, including their host galaxy demographics and spatial locations (motivated by observations of short GRBs, the BNS merger GW170817, and Type Ia SNe; $\S\ref{sec:demo}$), their rates ($\S\ref{sec:rates}$), lifetimes, dispersion and rotation measure, and the properties of their accompanying persistent radio sources ($\S\ref{sec:properties}$). Throughout the paper we compare the results to the properties of FRB~180924.

\section{Host Galaxies and Offsets}
\label{sec:demo}

\begin{figure*}
\centering
\includegraphics[width=0.95\textwidth]{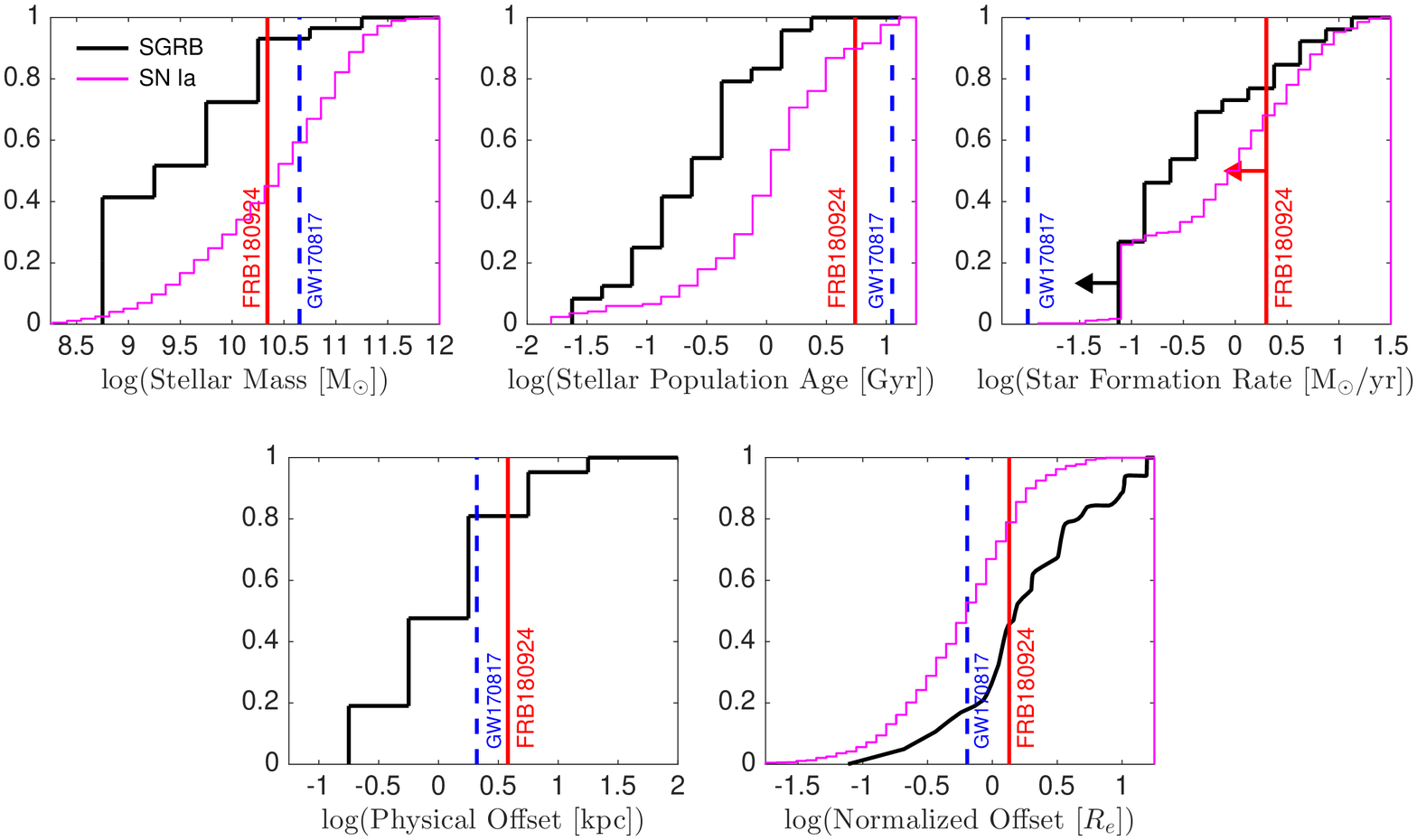}
\hspace{0.0cm}
\caption{Comparison of the host galaxy properties and offset of FRB~180924 (red; \citealt{Bannister+19}) to the population of SGRBs (black; \citealt{Berger14,Fong+17}), the BNS merger GW170817 (blue; \citealt{Blanchard+17,Fong+17}), and the hosts of Type Ia SNe (pink; \citealt{Neill+09,Gupta+11,Smith+12,Uddin+17}).  We show stellar mass, stellar population age, star formation rate, projected physical offset, and projected host-normalized offset (top-left to bottom-right).}
\label{fig:host}
\end{figure*}

FRBs that originate from magnetars created in SLSNe and LGRBs are expected to preferentially reside in low-metallicity dwarf galaxies known to host these classes of explosions (e.g., \citealt{Fruchter+06,Stanek+06,Modjaz+08,Lunnan+14,Perley+16,Schulze+18}).  Moreover, both LGRBs and SLSNe tend to concentrate in bright UV regions of their hosts \citep{Fruchter+06,Lunnan+15,Blanchard+16}, with a preference for small spatial offsets from their host centers \citep{Bloom+03,Lunnan+15,Blanchard+16}.  We have previously shown that the host (and location) of FRB~121102 are comparable to those of SLSNe and LGRBs \citep{Metzger+17,Nicholl+17c}.  The host galaxy of FRB~180924, on the other hand, does not match these expectations.

In contrast, BNS mergers are expected to occur in all types of galaxies (e.g., \citealt{Belczynski+06}), due to the broad delay time required for GW driven mergers. Furthermore, due to the natal kicks received by NSs at birth, BNS mergers can take place in locations spatially offset from their birth sites, sometimes outside of their host galaxies (e.g., \citealt{Fryer&Kalogera97,Bloom+99}).  The consistency of these predictions with the observed host galaxy demographics and spatial offset distributions of SGRBs provided strong evidence for the association of SGRBs with BNS mergers \citep{Berger14}, even prior to the discovery of GW170817 \citep{LIGO+17DISCOVERY,LIGO+17GRB}.  Though we have no direct probes of white dwarf AIC, their formation channels are thought to be sufficiently similar to Type Ia SNe that we might expect a similar host galaxy distribution.

In Figure~\ref{fig:host} we compare the host galaxy properties and spatial offset of FRB~180924 \citep{Bannister+19} to those of SGRBs \citep{Fong+13,Berger14,DePasquale19} and the BNS merger GW170817 \citep{Blanchard+17,Fong+17,Im+17}.  We also make a comparison to the host galaxies of Type Ia SNe \citep{Neill+09,Gupta+11,Smith+12,Uddin+17}, taking this as a proxy for the AIC scenario.  We find that in terms of stellar mass, star formation rate, and stellar population age, the host of FRB~180924 is comparable to those expected for BNS mergers and AIC.  The same is true for the physical and host-normalized offset of FRB~180924 relative to the distribution for SGBRs, GW170817, and Type Ia SNe.  Thus, from the point of view of its large-scale environment it is conceivable that FRB~180924 represents a BNS merger or AIC origin.  We stress that a more definitive statement will require a larger sample of localized FRBs \citep{Nicholl+17c}, as well as a broader comparison set of BNS mergers from GW detectors and from the long-sought class of AIC transients from future optical or radio surveys.

\section{Rates}
\label{sec:rates}

The observed SGRB rate \citep{Wanderman&Piran15} and an estimated beaming factor of $\sim 30$ \citep{Fong+15} leads to an estimate of the BNS merger rate at $z\lesssim 0.5$ of $R_{\rm BNS}\sim 300$ Gpc$^{-3}$ yr$^{-1}$.  This is consistent with the local merger rate estimated by Advanced LIGO/Virgo from the O1 and O2 observing runs of $110\lesssim R_{\rm BNS}\lesssim 3840$ Gpc$^{-3}$ yr$^{-1}$ \citep{LIGO+18CATALOG}, as well as the range required for BNS mergers to be the main source of heavy $r$-process elements in the Milky Way given the $r$-process yield inferred from the optical/infrared counterpart of GW170817 (e.g., \citealt{Cowperthwaite+17,Hotokezaka+18}).

It was proposed that during the final stages of a BNS merger inspiral the interaction between the NS magnetospheres could give rise to a single (non-repeating) FRB \citep{Hansen&Lyutikov01,Totani13,Zhang14,Wang+16,Metzger&Zivancev16}.  However, given the high volumetric rate of FRBs relative to BNS mergers, such ``one off'' bursts can account for at most only a small fraction, $\lesssim 1\%$, of the total FRB population \citep{Nicholl+17}. 

If, on the other hand, FRBs are produced well after the merger by a remnant magnetar, then each magnetar produced through this channel will produce many FRBs \citep{Nicholl+17,Yamasaki+18}.  However, as discussed in the next section, the magnetar remnant is initially enshrouded within the kilonova ejecta, which must become optically-thin to free-free absorption before FRBs can escape to an external observer, thus requiring a minimum magnetar lifetime of weeks to months. Meta-stable (hyper-massive or supra-massive) NS remnants will generally collapse to a black hole much earlier (e.g., \citealt{Shibata&Taniguchi06,Ravi&Lasky14}), unless the dipole magnetic field of the remnant is extremely weak, $\lesssim 10^{13}$ G, to prevent spin-down, despite the much larger fields generated during the merger process.  The mergers of binaries giving rise to indefinitely stable magnetar remnants, with masses at or below $M_{\rm TOV}$, are therefore the most promising FRB sources.

The fraction of BNS mergers that lead to stable remnants depends on the mass distribution of merging NS systems and the uncertain nuclear equation of state \citep{Piro+17}. \citet{Margalit&Metzger19} show that, for NSs drawn from the Galactic BNS population, and given current constraints on $M_{\rm TOV}$, at most $\approx 3\%$ of BNS mergers can produce stable NS remnants.    A low volumetric rate of stable NS remnants is also consistent with the lack of discovery of the time-evolving remnants of such objects in radio transient surveys \citep{Metzger+15b}, or in late-time follow up of SGRBs \citep{Fong+16}. Thus, the birth rate of magnetars capable of producing FRBs in the BNS merger channel is $\mathcal{R} \lesssim 0.03\mathcal{R}_{\rm BNS}\sim 3-100 \, {\rm Gpc}^{-1} \, {\rm yr}^{-1}$, comparable or less than the volumetric rate of LGRBs and SLSNe (e.g., \citealt{Prajs+17}), which are considered progenitors of FRBs with dwarf-galaxy hosts like FRB~121102 \citep{Tendulkar+17,Metzger+17,Nicholl+17c,Eftekhari+19}.  \citet{Nicholl+17} show that even a volumetric birth rate $\mathcal{R}\sim 10-100$ Gpc$^{-3}$ yr$^{-1}$ is sufficient to explain a significant portion of the observed FRB population if each magnetar is a similarly active FRB source for decades.  As discussed below (Equation~\ref{eq:tmag}), the predicted FRB active lifetime of massive magnetars produced in BNS mergers is consistent with this range.  

As there is no direct observational evidence for AIC, its rate remains highly uncertain.  Theoretical estimates of the AIC rate are in a range comparable to that of BNS mergers but also with uncertainties of over an order of magnitude (e.g., \citealt{Yungelson&Livio98,Tauris+13,Kwiatkowski15}).  The fraction of AIC events giving rise to magnetars via flux-freezing of the primordial white dwarf magnetic field is likely smaller than the total because of the low fraction, $\lesssim 15\%$, of strongly-magnetized white dwarfs (e.g., \citealt{Liebert+03}).  On the other hand, if the progenitor white dwarf is rapidly spinning, then even an initially weak magnetic field could be amplified to magnetar strengths following AIC by strong and differential rotation in the millisecond proto-NS remnant (e.g., \citealt{Dessart+07}).  Within the significant uncertainties, the magnetar birth rate via the AIC channel could therefore be comparable to that the BNS merger channel.

\section{FRB Properties}
\label{sec:properties}

\subsection{Bursts and Active Lifetime}
\label{sec:lifetime}

The origin of the coherent emission process responsible for FRB emission is uncertain.  One of the more developed models postulates that FRBs result from synchrotron maser emission from magnetized relativistic shocks \citep{Lyubarsky14,Beloborodov17}.  In this case each individual burst could be produced by the collision between an ultra-relativistic flare from the magnetar and a baryon-loaded shell released from an earlier flare \citep{Metzger+19}.  The external environment surrounding the magnetar on larger ($\sim$parsec) scales therefore does not directly impact the burst properties in this scenario. Nevertheless, variation in the average burst properties could still result from intrinsic differences in the magnetar's activity, as may be influenced by its mass, age, and magnetic field strength.

The mass distribution of Galactic BNS systems peaks above $M_{\rm TOV}$ \citep{Margalit&Metzger19}.  If representative of the extragalactic population, this implies that only a small fraction of mergers will leave stable NS remnants, and those will have masses just below $M_{\rm TOV}$.  Such extreme NSs may possess sufficiently high central densities to activate direct Urca cooling in their cores \citep{Page&Applegate92}, which exceeds by orders of magnitude the cooling rate via the ``modified'' Urca process that is believed to dominate in the majority of (less massive) NSs.  Because the rate of ambipolar diffusion of the magnetic field inside the NS depends sensitively on the core temperature \citep{BeloborodovLi16}, the timescale over which magnetic energy escapes from the stellar surface (e.g., in the form of FRB-generating flares) could speed up relative to magnetars born from core-collapse SNe or AIC.  Following \cite{BeloborodovLi16}, we estimate the magnetic activity timescale in the direct Urca (high-mass NS) and modified Urca (normal-mass NS) case as
\begin{equation}
t_{\rm mag} \sim 
\begin{cases}
20 \, {\rm yr} \, B_{16}^{-1} L_5^{3/2},  ~\text{high-mass NS}
\\
700 \, {\rm yr} \, B_{16}^{-1.2} L_5^{1.6}, ~\text{normal-mass NS}
\end{cases}
,
\label{eq:tmag}
\end{equation}
where we assume $\delta B\sim B/2$ as the amplitude of magnetic field fluctuations over a length-scale $L = 10^{5}L_{5} \, {\rm cm}$ inside the magnetar core.  The magnetic energy of the magnetar, $E_{\rm mag} \sim 3 \times 10^{49} \, B_{16}^2\, {\rm erg}$, therefore implies a characteristic dissipation power
\begin{equation}
\dot{E}_{\rm mag} \sim 
\frac{E_{\rm mag}}{t_{\rm mag}} \sim
\begin{cases}
5 \times 10^{40}  B_{16}^3 \, {\rm erg \, s}^{-1} \, ,  ~\text{high-mass NS}
\\
10^{39} B_{16}^{3.2} \, {\rm erg \, s}^{-1} \,  , ~\text{normal-mass NS}
\end{cases}
.
\label{eq:Edotmag}
\end{equation}
The most massive and highly magnetized magnetars born in BNS mergers may therefore be differentiated from lower-mass more weakly magnetized NSs such as those that might be created in SLSNe, LGRBs or AIC through their vastly different power output.

One may speculate that the shorter active lifetime of magnetars from BNS mergers could lead to shorter intervals between major ion ejection events relative to less massive magnetars, and an increased mass-loss rate.  In the shock-powered maser scenario, a larger average density surrounding the magnetar into which flare ejecta collides acts to increase the peak frequency of the bursts (Equation 47 of \citealt{Metzger+19}), possibly to values $\gg 1$ GHz, much higher than the sensitive range of FRB telescopes.  Lower, $\sim$GHz frequency bursts of the more readily-detectable kind might occur only in relatively rare cases in which magnetar activity ceases long enough to clear a low-density cavity around the magnetar.  Similar qualitative behavior (suppressed FRB emission for extended periods of time after major ion flares) was invoked to explain the long periods of inactivity observed for FRB~121102 \citep{Metzger+19}.  If the magnetars from BNS mergers undergo even longer periods of {\it apparent} inactivity at $\sim 1$ GHz, this could help explain why repeating bursts from FRB~180924 have not been detected despite significant followup \citep{Bannister+19}.  

Unlike in BNS mergers, the NS remnants of AIC would possess very low masses and would not be expected to undergo direct Urca cooling.  Nevertheless, one cannot exclude systematically different internal magnetic field strength or topology in magnetars formed from AIC versus those from core-collapse events, which could lead to qualitatively different bursting activity than the SLSNe/LGRB case resulting in different FRB properties.

\subsection{Ejecta Radio Transparency, Afterglow, and Dispersion Measure}
\label{sec:ejecta}

BNS mergers are accompanied by the ejection of radioactive neutron-rich material, which powers their optical/IR emission (e.g., \citealt{Li&Paczynski98}).  Axisymmetric hydrodynamical simulations find that the AIC of a slowly rotating, weakly magnetized white dwarf is accompanied by only a weak explosion that ejects a small amount of mass, $\lesssim 10^{-2}$ M$_{\odot}$, with a small fraction of radioactive $^{56}$Ni (e.g., \citealt{Dessart+06,Abdikamalov+10}).  However, if the white dwarf is strongly magnetized and rapidly rotating prior to collapse (the type of events most likely to give birth to millisecond magnetars), then the amount of unbound ejecta can be substantially larger, $\sim 0.01-0.1$ M$_{\odot}$, closer to that of BNS mergers \citep{Dessart+07,Metzger+09b}.  

The expanding ejecta shell surrounding the nascent magnetar in the BNS merger/AIC scenarios affects their observational signatures.  First, to be observable an FRB must propagate through this confining medium, which is initially optically thick to free-free absorption at $\sim$GHz frequencies. The free-free optical depth is set by the ejecta density, temperature, and ionization state. Here we calculate the free-free optical depth of homologously expanding BNS merger/AIC ejecta for which the temperature and ionization state are governed by photo-ionization from spin-down of the nascent magnetar. We use the publicly available photo-ionization code CLOUDY \citep{Ferland+13} and follow the methods described in \citet{Margalit+18} for magnetars embedded in SLSNe ejecta.

\begin{figure}
\includegraphics[width=0.5\textwidth]{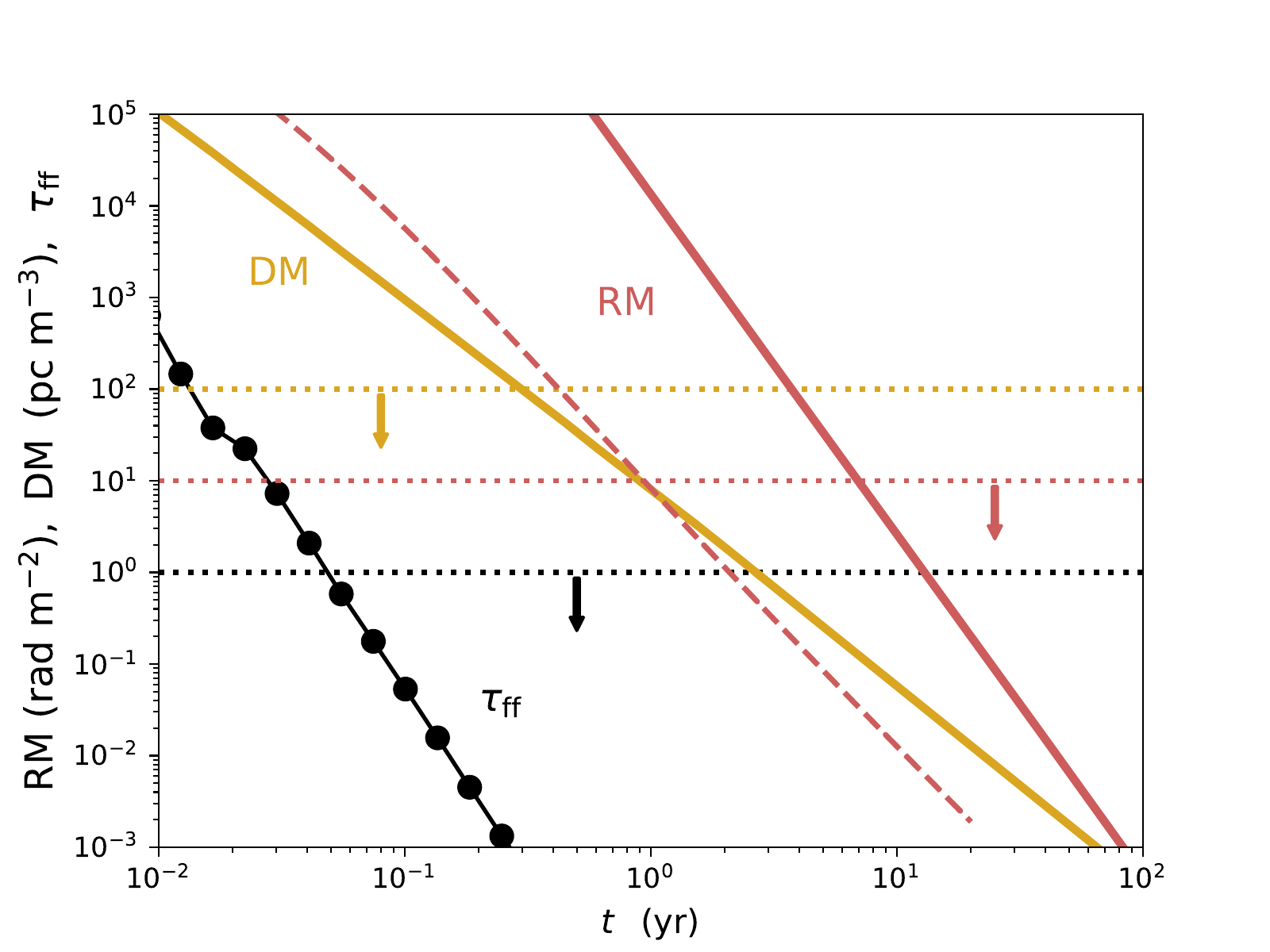}
\hspace{0in}
\caption{Temporal evolution of the free-free optical depth at $1.4$ GHz (black) and of the dispersion measure (yellow) imposed by the BNS merger/AIC ejecta, as calculated using the photo-ionization code CLOUDY.  Also shown is the rotation measure (red) arising from the magnetar nebula for two representative models described in the text: model A (solid) and model B (dashed). Horizontal dotted curves show upper limits on the various quantities for FRB~180924 \citep{Bannister+19}.}
\label{fig:nebula}
\end{figure}

A primary difference between the SLSN/LGRB and BNS merger/AIC scenarios is the lower ejecta mass and higher ejecta velocities in the latter case, which results in a shorter free-free transparency time, $t_{\rm ff}$. Naively, for a fixed ionization fraction one expects $t_{\rm ff} \propto M_{\rm ej}^{2/5} v_{\rm ej}^{-1}$, in which case $t_{\rm ff}$ will be $\sim 100$ times shorter for the BNS merger/AIC case, i.e. months instead of $\sim$decade timescale for SLSNe \citep{Margalit+18}. However, even this overestimates the free-free transparency timescale by a factor of a few since it does not account for temperature effects on the ionization state.  The black curve in Figure~\ref{fig:nebula} shows the result of our CLOUDY calculations for the ionization state of BNS merger/AIC ejecta, calculated assuming a central ionizing source equal to the spin-down luminosity of the magnetar for assumed dipole magnetic field strength $B\sim 10^{14}-10^{16} \, {\rm G}$ and birth period $\sim 1$ ms.  At 1.4~GHz we find a transparency timescale of $t_{\rm ff}\sim 1$~week for an assumed ejecta mass $M_{\rm ej} = 0.1 M_\odot$ and velocity $v_{\rm ej} = 0.52c$ (Equation~\ref{eq:vej}); this high value of $v_{\rm ej}$ results if the ejecta is accelerated by the spin-down energy of the magnetar early in its evolution, when radiation is still trapped and goes into $PdV$ expansion (see Equation~\ref{eq:vej} below).  A more conservative upper bound on $t_{\rm ff}$ is obtained by neglecting magnetar acceleration; adopting the ejecta mass and velocity inferred for GW170817 by \citet{Villar+17}, we find $t_{\rm ff}\sim {\rm month}$ at 1.4~GHz, with the detailed result depending on the ionizing radiation field (magnetic dipole field strength).

In the BNS merger/AIC scenario, the rapidly diluting ejecta becomes nearly fully ionized by the magnetar radiation. The contribution to the burst DM by the ejecta, which can be extracted directly the CLOUDY simulations (yellow curve in Figure~\ref{fig:nebula}), is therefore easily approximated by
\begin{equation}
    {\rm DM}_{\rm ej} \approx \frac{3 M_{\rm ej}}{8 \pi m_{\rm p} (v_{\rm ej} t)^2} \approx 5 \, {\rm pc \, cm}^{-3} \, M_{{\rm ej},-1} \beta_{\rm ej}^{-2} t_{\rm yr}^{-2},
\end{equation}
where $\beta_{\rm ej} = v_{\rm ej}/c$ and we have assumed a fully ionized heavy ejecta composition (i.e., atomic mass $A\approx 2Z$).

\begin{figure*}
\centering
\includegraphics[width=0.9\textwidth]{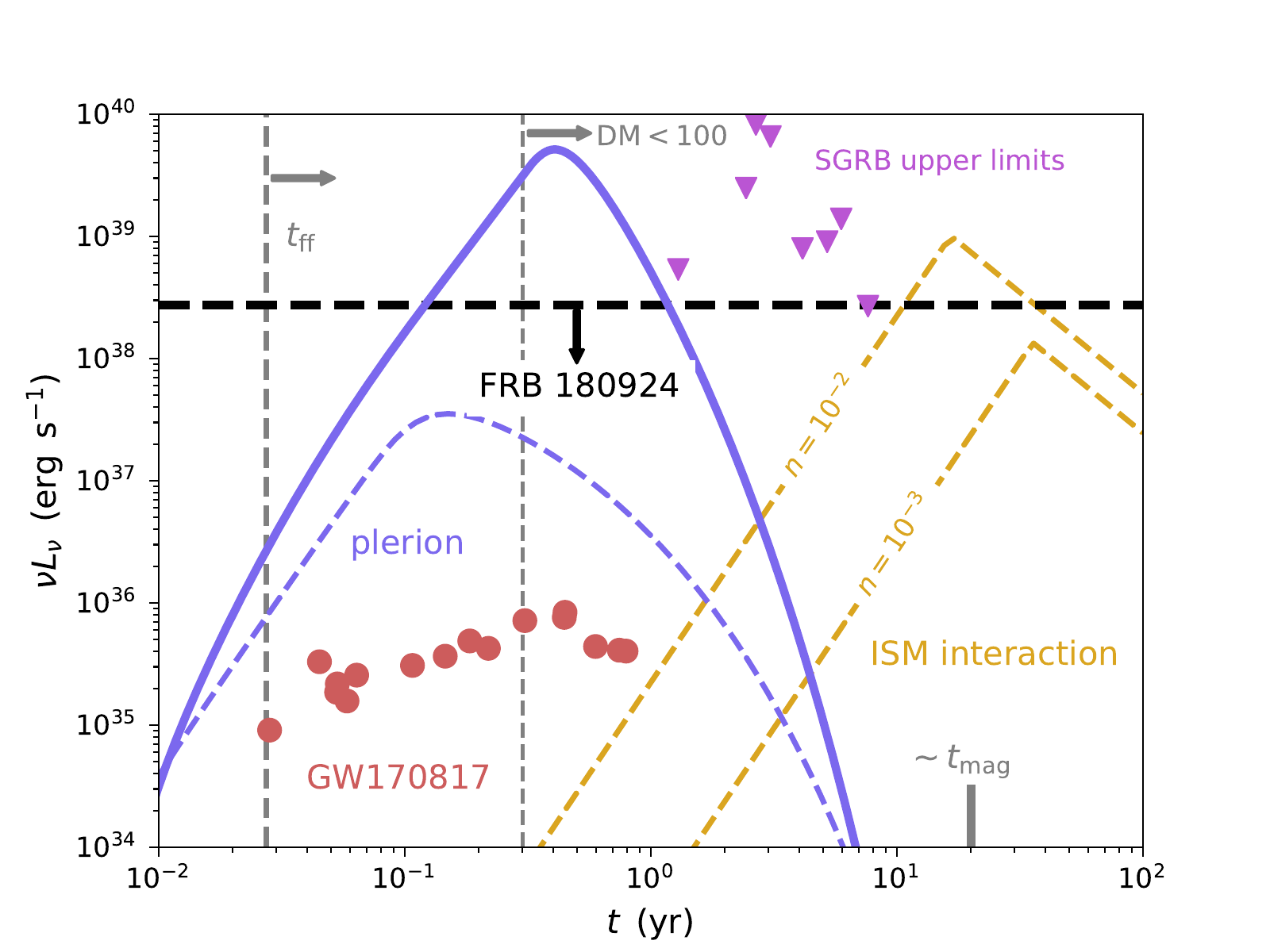}
\caption{Potential sources of long-lasting persistent synchrotron radio emission associated with FRBs generated by magnetars born in BNS mergers/AIC.  These include plerionic emission from a magnetar-wind-inflated nebula behind the merger/AIC ejecta (blue curves show two example models, as in Figure~\ref{fig:nebula}), and the afterglow as the ejecta interacts with the ISM (yellow dashed curves; ISM densities of $n=10^{-2}$ and $10^{-3} \, {\rm cm}^{-3}$).  Also shown is the upper limit on persistent radio emission associated with FRB~180924 (horizontal dashed curve), the free-free transparency time, $t_{\rm ff}$, of the merger ejecta, the time at which the ejecta DM decreases to $\lesssim 100 \, {\rm pc \, cm}^{-3}$, the observed jet-afterglow emission for GW170817 (red circles; \citealt{Alexander+18}), and upper limits from SGRBs (purple triangles; \citealt{Fong+16}).}
\label{fig:radio}
\end{figure*}

In addition to generating free-free absorption and local DM at early epochs, the BNS merger/AIC ejecta will produce a late-time ($\sim$decade long) radio afterglow from its interaction with the ambient interstellar (e.g., \citealt{Nakar&Piran11}) or circumstellar (e.g., \citealt{Moriya16}) medium.  This radio signature may be particularly prominent in the scenario considered here, since formation of a long-lived magnetar remnant and its subsequent spin-down are likely to inject an enormous amount of rotational energy of up to $\sim 10^{53} \, {\rm erg}$ into the surrounding medium \citep{Metzger&Bower14}.  In most cases in which the dipole spin-down timescale is much shorter than the radiative diffusion time, this will accelerate the BNS merger/AIC ejecta to high velocities, 
\begin{equation}
    \beta_{\rm ej} = \sqrt{1 - \left( 1 + {E}/{M_{\rm ej}c^2} \right)^{-2}} \approx 0.52 \, E_{52.5}^{1/2} M_{{\rm ej},-1}^{-1/2},
    \label{eq:vej}
\end{equation}
and induce a significantly stronger late-time radio signature than for BNS mergers that do not form such remnants; for GW170817, the total kinetic energy of the kilonova ejecta is estimated to be significantly smaller, $\approx 2.5 \times 10^{51}$ erg \citep{Villar+17}, consistent with the arguments that GW170817 did not form a long-lived magnetar \citep{Margalit&Metzger17}. Additionally, searches for enhanced late-time radio emission following SGRBs have placed limits on the fraction of BNS mergers that form magnetar remnants \citep{Metzger&Bower14,Fong+16,Horesh+16}, consistent with the estimates adopted in \S\ref{sec:rates}.

The kinetic energy transferred to the ejecta via magnetic-dipole spin-down may be smaller than the total rotational energy of the magnetar due to inefficiencies in coupling the magnetar wind to the BNS merger/AIC ejecta (e.g., \citealt{Bucciantini+12}), and hence we adopt $E = 3\times 10^{52} \, {\rm erg}$ as a fiducial value\footnote{Even this energy deposition would be overestimated if the NS is sufficiently deformed into a non-axisymmetric shape such that that GW spin-down dominates over magnetic-dipole spin-down \citep{Ai+18}; however, this requires extreme NS ellipticities, which may not be sustainable given that the required magnetic field configurations are not MHD stable except in a narrow region of parameter space \citep{Margalit&Metzger17}.} in Equation~\ref{eq:vej}. The dashed yellow lines in Figure~\ref{fig:radio} show a simplified model for the radio light curve produced by the interaction of the magnetar-accelerated ejecta with the ISM, using the formulation of \citet{Nakar&Piran11} and assuming $M_{\rm ej} = 0.1 M_\odot$, $E = 3 \times 10^{52} \, {\rm erg}$, $v_{\rm ej}$ given by Equation~\ref{eq:vej}, and different ISM densities.  The blue lines in Figure~\ref{fig:radio} also show a model for magnetar plerion emission described in the next section.  
For comparison, we also show the upper limit on persistent radio emission at the location of FRB~180924 \citep{Bannister+19}, the timescale for the ejecta to become transparent to free-free absorption at $1.4$ GHz, and the time after which the ejecta DM decreases to $\lesssim 100 \, {\rm pc \, cm}^{-3}$, consistent with upper limits on the local DM contribution of FRB~180924 \citep{Bannister+19}.  Although it is unclear whether a relativistic jet is launched from long-lived magnetars in BNS mergers (e.g., \citealt{Murguia-Berthier+14}), we also show for comparison the measured jet radio afterglow emission of GW170817 \citep{Alexander+18}.  Even were radio emission akin to GW170817 produced in magnetar-remnant scenarios, it would not be sufficiently luminous to be detected at the distances of FRBs such as 180924.

Interaction of the magnetar-accelerated ejecta with the ISM may or may not contribute detectable radio emission, depending sensitively on the ISM density. SGRB afterglow modeling suggests small typical densities, $n\sim 10^{-2}$ cm$^{-3}$, consistent with the typical large host-galaxy offsets \citep{Fong+15}.  The local ISM density in the vicinity of FRB~180924 may be similarly small, in agreement with the implied low local DM contribution and the location on the outskirts of its host galaxy.  We caution that the ejecta light curves in Figure~\ref{fig:radio} are highly simplified; deviations from spherical symmetry will generally delay the peak timescale \citep{MargalitPiran15}, while emission earlier than the Sedov-Taylor time of the ejecta (at which the emission peaks) is extremely sensitive to the high-velocity tail of the ejecta distribution (e.g., \citealt{HotokezakaPiran15}).  

\subsection{Nebula Persistent Radio Emission and RM}
\label{sec:nebula}

The large RM and persistent radio flux of FRB~121102 are both consistent with arising from a nebula of magnetic fields and electrons injected by the magnetar flares and confined by the SN ejecta \citep{Beloborodov17,Margalit&Metzger18}.  A similar idea for rotationally-powered plerionic emission from AIC and FRB sources was discussed by \citet{Piro&Kulkarni13} and \citet{Murase+16}, respectively.  These characteristics depend both on intrinsic properties of the magnetar engine, such as its energy injection rate, history, and mass loading, but also on the age of the system and circum-engine environment. 

While it is not clear how the intrinsic properties of the FRB activity vary for magnetars born in different astrophysical settings (although see Equations~\ref{eq:tmag} and \ref{eq:Edotmag} and surrounding discussion), the external environment is more robustly predicted to differ markedly for magnetars formed in SLSNe/LGRBs versus those created in BNS mergers/AIC.  In the former, the magnetar nebula expands within the dense SN ejecta of $M_{\rm ej} \sim$several $M_\odot$ and $v_{\rm ej}\sim 10^4 \, {\rm km \, s}^{-1}$, while in the BNS merger/AIC case the confining medium within which the nebula expands is the faster, more dilute BNS merger/AIC ejecta ($M_{\rm ej}\sim 0.1 M_\odot$ and $v_{\rm ej}\sim 0.5 c$; Equation~\ref{eq:vej}). This allows the nebula to expand at a faster rate in the BNS merger/AIC scenario and leads to a more rapid decrease in the nebula density and magnetic field, and hence also in the imprinted RM and accompanying plerionic radio emission. 

The velocity of the nebula inflated within a homologously expanding uniform-density ejecta can be roughly estimated as
\begin{equation}
    v_{\rm n} \sim \left( \frac{ \int \dot{E}_{\rm n} dt \, v_{\rm ej}^3}{M_{\rm ej}} \right)^{1/5} \approx 0.2 c \, E_{{\rm n},50}^{1/5} \beta_{\rm ej}^{3/5} M_{{\rm ej},-1}^{-1/5} ,
    \label{eq:vn}
\end{equation}
where we have adopted fiducial values for instantaneous (present-day) nebula energy $E_{\rm n}$ and the BNS merger ejecta.  For ejecta parameters appropriate to SLSNe, Equation~\ref{eq:vn} predicts nebula velocities smaller by an order of magnitude, $\sim 4,000 \, {\rm km \, s}^{-1}$, consistent with $v_{\rm n}$ adopted for the models of FRB~121102 in \citet{Margalit&Metzger18}.  To account for the lower ejecta density in a more detailed manner, we self-consistently integrate the dynamical equations for the nebula's time-dependent evolution in the thin-shell approximation as energy is deposited by the central engine (e.g., \citealt{Chevalier05}). This is a direct extension of the \citet{Margalit&Metzger18} model, which assumed constant $v_{\rm n}$ for simplicity.

In Figure~\ref{fig:nebula} we show the resulting nebula radio light curve and RM for two models. Model A (solid blue curve) is a scaled version of the model fit to the observed persistent source of FRB~121102 (Model A of \citealt{Margalit&Metzger18}), where the characteristic timescale of magnetar energy deposition has been reduced by a factor of $\sim 35$, in line with the shorter estimated value of $t_{\rm mag}$ (Equation~\ref{eq:tmag}). Additionally, as discussed above, the nebula expansion velocity within the ejecta is explicitly integrated as a function of time (as opposed to being a constant free parameter as in \citealt{Margalit&Metzger18}), and the ejecta parameters adopted are $M_{\rm ej}=0.1M_\odot$, $v_{\rm ej}=0.52 c$ (Equations~\ref{eq:vej},\ref{eq:vn}).  We find that the DM contributed by the nebula is always sub-dominant to that imprinted by the BNS merger/AIC ejecta, and can thus be neglected. The scaled model of FRB~121102 is consistent with both the lack of persistent radio emission and the low measured RM of FRB~180924 \citep{Bannister+19} for a source age of $\gtrsim 10$ yr following a BNS merger/AIC.

We also explore a second model (Model B; dashed blue curve in Figure~\ref{fig:radio}) in which the magnetar engine deposits a total energy of $E_{\rm mag} = 3 \times 10^{49} \, {\rm erg}$ at a constant rate over a timescale $t_{\rm mag} = 20 \, {\rm yr}$.  These are plausible parameters for a magnetar with an internal magnetic field strength $B_\star = 10^{16} \, {\rm G}$ given the active lifetime predicted by ambipolar diffusion\footnote{However, note that \citet{Margalit&Metzger18} found that a constant energy-injection-rate model cannot simultaneously reproduce both the RM and radio luminosity for FRB~121102.} (Equation~\ref{eq:tmag}).  Figures~\ref{fig:nebula},\ref{fig:radio} show that for this model, the upper limits on RM and persistent radio flux of FRB~180924 are satisfied by even younger magnetars, $\gtrsim 1$ yr old.

\section{Archival Search for a Short GRB Coincident with FRB~180924}

Motivated by the possible origin of FRB~180924 in a BNS merger, we searched the BATSE, {\it Fermi}/GBM, and {\it Swift}/BAT catalogs for SGRBs consistent with the FRB position.  We found no such sources from {\it Swift}/BAT, which provides the best positional accuracy of all three detectors (typically $3'$ radius), but has an instantaneous sky coverage of only about $15\%$.

In the BATSE catalog, which spans April 1991 to May 2000, with full sky instantaneous coverage, two SGRBs (921115 and 940114) are consistent with the location of FRB~180924 to within about 1.5 times their large error radii of $12.7^\circ$ and $9^\circ$, respectively. These error regions corresponds to about $1.4-2.8\%$ of the sky, so given the sample of about 500 SGRBs in the BATSE catalog, it is not surprising that two SGRBs would overlap the location of FRB~180924 by pure coincidence.

Similarly, in the {\it Fermi}/GBM catalog, which extends from July 2008 to the present and provides an instantaneous sky coverage of about $65\%$, we identify 5 SGRBs that overlap the location of FRB~180924 (GRBs 121014, 131006, 141128, 150805, and 170125), but again with large error radii of $9-27^\circ$ (and nominal offsets of $0.2-1$ times the GBM error radius); these error circles correspond to about $0.6-5.4\%$ of the sky.  Given a sample of 420 SGRBS in the GBM catalog, it is therefore not surprising to find several events that overlap the location of FRB~180924.

Thus, we cannot conclusively link FRB~180924 to a specific SGRB, but plausible candidates exist in both the BATSE and {\it Fermi}/GBM catalogs.

\begin{figure*}
\begin{center}
\includegraphics[width=0.95\textwidth]{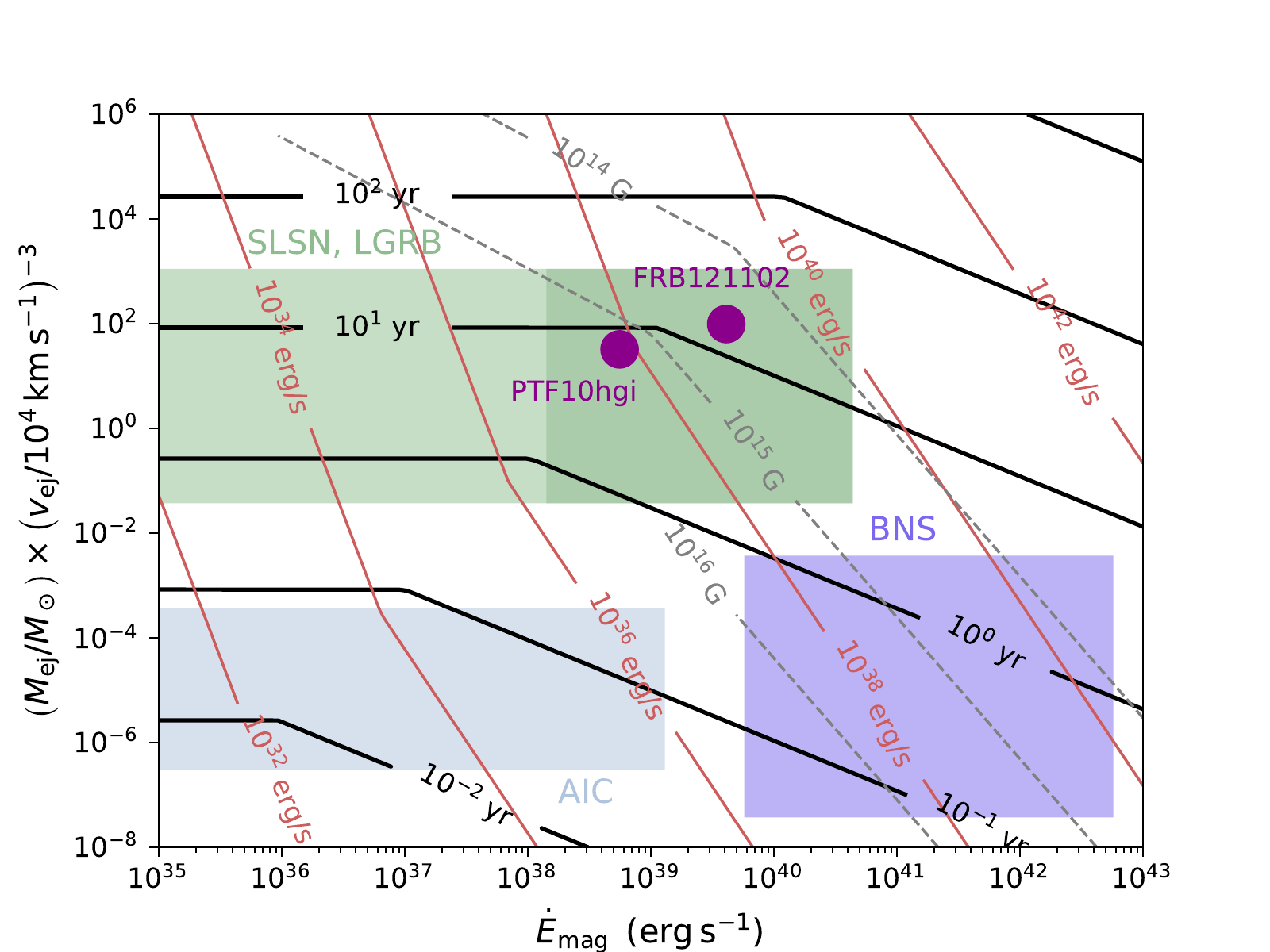}
\end{center}
\hspace{0.1in}
\vspace{-0.2in}
\caption{Properties of the persistent nebular radio emission from magnetars formed from different progenitor channels (SLSNe, LGRBs, AIC, and BNS mergers) in the space of intrinsic and extrinsic properties.  Red and black contours show the peak luminosity, $\nu L_{\nu}$, and peak timescale of the radio light-curve at $6 \, {\rm GHz}$, respectively (Equation~\ref{eq:peak_radio}).  The peak timescale is set by the shorter of that required for the nebula to become transparent to synchrotron self-absorption and for the ejecta to become transparent to free-free absorption (right or left of the break in the contours, respectively).  These properties differ based on the (assumed constant) energy injection rate, $\dot{E}_{\rm mag}$ (horizontal axis; Equation~\ref{eq:Edotmag}), a proxy for the internal magnetar $B$ field strength (Equations~\ref{eq:tmag} and \ref{eq:Edotmag}), and on the ejecta density as parameterized by $M_{\rm ej}/v_{\rm ej}^{3}$ (vertical axis), where $M_{\rm ej}$ and $v_{\rm ej}$ are the ejecta mass and mean velocity, respectively.  Gray dashed lines show the {\it external dipole} magnetic field strength below which the spin-down luminosity exceeds the magnetic power, $\dot{E}_{\rm mag}$.  Also shown are the radio sources coincident with FRB~121102 \citep{Chatterjee+17} and the SLSN PTF10hgi \citep{Eftekhari+19}.}
\label{fig:parameterspace}
\end{figure*}

\section{Discussion and Conclusions}

Motivated by the localization of FRB~180924 to the outskirts of a massive early-type host galaxy \citep{Bannister+19}, we explore and develop a model for FRBs arising from magnetars born in BNS mergers and/or AIC \citep{Nicholl+17,Yamasaki+18}. The environment of FRB~180924 is dramatically different from that of the repeating FRB~121102, whose low-metallicity star forming dwarf host galaxy first led to the suggested association of FRBs with magnetars born in SLSNe and/or LGRBs \citep{Metzger+17,Nicholl+17}.  Here we show that the host galaxy and offset of FRB~180924 are well-matched to the distributions for SGRBs and Type Ia SNe, which are proxies for BNS mergers and AIC events, respectively. We search archival data for a possible SGRB coincident with the location of FRB~180924 and find plausible candidates, although a definitive association cannot be made.

We demonstrate several likely differences between the BNS merger, AIC, and SLSN/LGRB magnetar birth scenarios and how these may affect observable properties of FRBs produced by such magnetars. These generally separate into intrinsic differences in the magnetar source activity, and external differences in the surrounding environment.  Although the intrinsic differences are rather uncertain, we raise the possibility that stable magnetars born in BNS mergers would form the most massive NSs and may thus cool more efficiently via direct Urca cooling than standard lower mass NSs.  This leads to a shorter ambipolar diffusion timescales and faster extraction of the magnetar's magnetic energy (Equations~\ref{eq:tmag} and \ref{eq:Edotmag}; see \citealt{BeloborodovLi16}). Extrinsic differences in the ambient medium are clearer: the ejecta into which the magnetar is born is more dilute in both the BNS merger and AIC scenarios in comparison to the SLSN/LGRB channel. The density of this ambient medium governs the expansion velocity of the magnetar nebula and therefore the timescale on which synchrotron emission from the deflating nebula peaks.

In Figure~\ref{fig:parameterspace} we show the parameter space of nebular synchrotron emission as a function of  intrinsic (horizontal axis) and extrinsic (vertical axis) properties. Black (red) contours show the peak time (luminosity) at $6 \, {\rm GHz}$ of the persistent radio source associated with the magnetar nebula as a function of the magnetar engine power, $\dot{E}_{\rm mag}$, and the surrounding ejecta density, parameterized via $M_{\rm ej} v_{\rm ej}^{-3}$. These two parameters fully describe the nebula in the simple case assumed here in which the rate of particle and energy injection into the nebula are constant in time.  Although adopted for simplicity of presentation and analytic tractability, we note that such a toy model was shown to not be able to reproduce in detail both the RM and persistent radio luminosity of FRB~121102 \citep{Margalit&Metzger18}.

Figure~\ref{fig:parameterspace} illustrates qualitative differences between magnetar nebulae in the different scenarios.  The shaded regions of different colors show expectations for the different channels (SLSLN, LGRB, AIC, BNS mergers) spanning low/high internal magnetar $B$-field strengths, with/without direct Urca cooling (Equation~\ref{eq:Edotmag}), and ejecta with both high and low mass/velocity.  As discussed previously, due to the high ejecta velocity and low ejecta mass, the nebular radio emission from magnetars formed in BNS mergers and/or AIC peaks at early times, as does the time at which the ejecta become transparent to free-free absorption (visible as kinks in the black contours in Figure~\ref{fig:parameterspace}).  This may render the persistent nebular radio emission of such sources more difficult to detect in targeted searches of known FRB positions than in the case of SLSNe/LGRBs, consistent with the non-detection of persistent emission from FRB~180924 in comparison with FRB~121102, as well as the lower RM of FRB~180924 (Figure~\ref{fig:nebula}).

The discovery of FRB emission from the remnants of either NS mergers or AIC would have major implications for our understanding of these still poorly-understood events.  Although frequently invoked, AIC has not yet been directly observed, and indeed even whether a O/Ne WD approaching the Chandrasekhar mass will undergo AIC or a thermonuclear detonation (producing a SN Ia-like transient) remains debated (e.g., \citealt{Jones+19}).  Likewise, the fraction of BNS mergers giving rise to long-lived stable remnants is extremely sensitive to uncertain properties of the NS equation-of-state \citep{Margalit&Metzger19}.  If a subset of FRBs originate from BNS merger remnants, their observed properties and environments relative to the comparatively unbiased merger population detected by LIGO/Virgo would offer unique insights into the diverse outcome of mergers and the properties of NSs.  
If FRBs are observed $\sim$weeks--decades following BNS mergers detected by LIGO/Virgo, this may provide an invaluable means of differentiating supra-massive from indefinitely-stable NS remnants and place new constraints on the NS equation-of-state \citep{Margalit&Metzger19}.

An additional signature that may be associated with FRBs from BNS mergers and/or AIC is late-time ($\sim$decade timescale) radio emission due to interaction of the dynamical merger ejecta with the ambient medium. If accelerated by the initial spin-down energy deposition of the newborn magnetar, this signal may be detectable even at large distances and is also a useful diagnostic in constraining the NS equation-of-state \citep{Margalit&Metzger17}.  As one test of the hypothesis that FRBs originate from a subset of BNS mergers that produce stable magnetar remnants, and as a future means of constraining the properties of NSs, we propose late-time searches for FRB emission from the locations of SGRBs and GW-detected BNS mergers.

\acknowledgements
BDM and BM acknowledge many conversations with Andrei Beloborodov. BM would like to thank the organizers and participants of FRB2019 for a highly enjoyable and productive meeting. BM is supported by NASA through the NASA Hubble Fellowship grant \#HST-HF2-51412.001-A awarded by the Space Telescope Science Institute, which is operated by the Association of Universities for Research in Astronomy, Inc., for NASA, under contract NAS5-26555. BDM acknowledges support from the NASA Astrophysics Theory Program (grant NNX16AB30G). The Berger Time-Domain Group is supported in part by NSF grant AST-1714498 and NASA grant NNX15AE50G.


\appendix
\section{Analytic Solution for Plerionic Radio Emission}
In the following we derive an approximate solution describing the nebular radio emission in the simplified case of temporally-constant energy (and particle) injection rate a magnetar embedded within a homologously expanding ejecta with a radially-constant density distribution.
This model is therefore described by the two parameters $\dot{E}$ and $\zeta \equiv M_{\rm ej}/v_{\rm ej}^3$, where $\zeta$ is a proxy for the ejecta density ($\rho_{\rm ej} \sim \zeta t^{-3}$).

If radiative losses are negligible and treating the nebula plasma as a relativistic fluid, then the nebula expands within the confining ejecta (which is itself assumed to be homologously expanding) with a velocity
\begin{equation}
v_{\rm n} \sim \left( \frac{\dot{E} t}{\zeta} \right)^{1/5}
\end{equation}
and corresponding size
\begin{equation}
R_{\rm n} = \int v_{\rm n} \, dt 
\sim \frac{5}{6} \left(\frac{\dot{E}}{\zeta}\right)^{1/5} t^{6/5} .
\end{equation}

Assuming a tangled nebular magnetic field, adiabatic losses imply the field strength is determined by the magnetic energy injected within the last dynamical (nebula-expansion) timescale, $\sim \sigma \dot{E}t$, where $\sigma$ is the ratio of magnetic to kinetic energy injected by the magnetar. The nebula magnetic field is therefore
\begin{equation} \label{eq:Bn}
B_{\rm n} \sim \left( \frac{6 \sigma \dot{E} t}{R_{\rm n}^3} \right)^{1/2} 
\sim \left(\frac{6}{5}\right)^{3/2} \left( 6 \sigma \right)^{1/2} \dot{E}^{1/5} \zeta^{3/10} t^{-13/10} .
\end{equation}

\subsection{Electron Distribution}

We assume electrons are injected into the nebula with a characteristic Lorentz factor $\gamma_{\rm inj}$ and at a constant rate,
\begin{equation} \label{eq:Ndot}
\dot{N}_\gamma\left(\gamma\right) \sim \chi^{-1} \dot{E} \delta \left( \gamma - \gamma_{\rm inj} \right) ,
\end{equation}
where $\gamma_{\rm inj} \approx \chi / 2 m_e c^2$ and $\chi \gtrsim 0.2 \, {\rm GeV}$ the mean energy per $e-p$ ejected in each baryon flare (the minimum value is set by the escape speed from the NS surface, and for $\gamma_{\rm inj}$ we assume the electrons and protons enter the nebula in equipartition after passing through the wind-termination shock).

Given the electron injection rate and some specified cooling process setting $d\gamma/dt = \dot{\gamma}_{\rm cool}(\gamma,t)$, the particle distribution can be calculated from number conservation, which implies\footnote{note that here $N_\gamma \equiv (\partial N / \partial \gamma)$ is the {\it total number} of electrons per unit Lorentz factor, whereas in some previous work \citep[e.g.][]{Margalit&Metzger18} this notation was used for particle {\it number density} per unit Lorentz factor instead).}
\begin{equation} \label{eq:distributionIntegral}
N_\gamma (\gamma,t) = \int_\gamma^{\infty} \dot{N}_\gamma(\gamma_0,t_0) \left(\frac{\partial t_0}{\partial \gamma}\right) \, d \gamma_0 .
\end{equation}
Here the subscript zero conveys that particles of Lorentz factor $\gamma$ at time $t$ had at earlier times $t_0 < t$ a higher Lorentz factor $\gamma_0 > \gamma$, and the relation between $\gamma,t,\gamma_0,t_0$ is determined by the cooling process.

For adiabatic cooling,
\begin{equation}
\dot{\gamma}_{\rm cool}(\gamma,t) = \dot{\gamma}_{\rm ad} = - \frac{\gamma}{t},
\end{equation}
and the solution to the differential equation $d\gamma/dt = \dot{\gamma}$ is trivial,
\begin{equation}
t_0 = t \gamma / \gamma_0 \,\,\, , \,\,\, {\rm adiabatic \, cooling} .
\end{equation}
The resulting distribution implied by Equation~(\ref{eq:distributionIntegral}) is then
\begin{equation}
N_\gamma (\gamma,t) = 
\frac{\dot{N}_{\rm i} t_{\rm i}}{\gamma_{\rm inj}} \left(\frac{t}{t_{\rm i}}\right)^{1-\alpha} \left(\frac{\gamma}{\gamma_{\rm inj}}\right)^{-\alpha}
= \frac{\dot{E}}{\chi \gamma_{\rm inj}}
\begin{cases}
0 &, \gamma > \gamma_{\rm inj}
\\
\gamma^0 t &, \gamma < \gamma_{\rm inj}
\end{cases}
\,\,\,\,\,\, , \,\,\, {\rm adiabatic \, cooling} .
\end{equation}
Here in the first equality we have generalized the particle injection (Equation~\ref{eq:Ndot}) to a power-law time dependence, $\dot{N}_\gamma = \dot{N}_{\rm i} (t / t_{\rm i} )^{-\alpha} \delta ( \gamma - \gamma_{\rm inj} )$, and in the last line we have reverted to the constant particle injection case considered here for simplicity ($\alpha=0$).

For synchrotron cooling with a general power-law decaying magnetic field $B_{\rm n} \propto t^{-\beta/2}$ (as in Equation~\ref{eq:Bn}),
\begin{equation}
\dot{\gamma}_{\rm cool}(\gamma,t) = \dot{\gamma}_{\rm syn} = - \gamma^2 \frac{\sigma_T c B_{\rm n}^2 / 8 \pi}{m_e c^2}
\equiv - \gamma^2 t^{-\beta} A ,
\end{equation}
and the solution for the Lorentz factor evolution with time implies
\begin{equation}
t_0 = \left[ \frac{\beta-1}{A} \left( \gamma^{-1} - \gamma_0^{-1} \right) + t^{-(\beta-1)} \right]^{-1/(\beta-1)}
\,\,\, , \,\,\, {\rm synchrotron \, cooling} .
\end{equation}
For the case of interest in this model (Equation~\ref{eq:Bn}), we have
\begin{equation}
\beta = 13/5
\,\,\, , \,\,\,
A = \frac{3 \sigma_{\rm T}}{4 \pi m_e c} \left(\frac{6}{5}\right)^{3} \sigma \dot{E}^{2/5} \zeta^{3/5} .
\end{equation}
The electron Lorentz factor distribution (Equation~\ref{eq:distributionIntegral}) can then be integrated to give
\begin{align} \label{eq:N_gamma_syn}
N_\gamma (\gamma,t) = 
\dot{N}_{\rm i} \left(\frac{t_0[\gamma,t]}{t_{\rm i}}\right)^{-\alpha} \frac{1}{A \gamma^2} \left[ \frac{\beta-1}{A} \left( \gamma^{-1} - \gamma_{\rm inj}^{-1} \right) + t^{-(\beta-1)} \right]^{-\beta/(\beta-1)}
\nonumber \\
\longrightarrow
\frac{\dot{E}}{A \chi}
\begin{cases}
0 &, \gamma > \gamma_{\rm inj}
\\
\gamma^{-2} t^{\beta} &, \gamma_{\rm cool}(t) \ll \gamma \ll \gamma_{\rm inj}
\\
\gamma^{\frac{2-\beta}{\beta-1}} t^0 \left(\frac{A}{\beta-1}\right)^{\frac{\beta}{\beta-1}} &, \gamma \ll \gamma_{\rm cool}(t)
\end{cases}
\,\,\,\,\,\, , \,\,\, {\rm synchrotron \, cooling},
\end{align}
where $\gamma_{\rm cool}(t) \equiv A^{-1} (\beta-1) t^{\beta-1}$
and in the last equality we have again reverted to the simplified case considered here where the particle injection rate is constant in time, i.e. $\alpha=0$.

\subsection{Synchrotron Radiation}

Optically thick synchrotron emission from the nebula is easy to calculate first as it does not depend on the particle distribution, which is the major complexity in analytically describing the nebula. We remember that the Lorentz factor of an electron whose peak emission is at frequency $\nu$, is
\begin{equation}
\gamma(\nu,t) \approx \left( \frac{2\pi m_e c \nu}{e B_{\rm n}(t)} \right)^{1/2} .
\end{equation}
The optically thick synchrotron emission can be approximated as $L_\nu \approx 8 \pi^2 R_{\rm n}^2 kT(\gamma) \nu^2/c^2$ with $kT(\gamma) \approx \gamma m_e c^2 / 3$, resulting in
\begin{align} \label{eq:Lnu_thick}
L_\nu \left( \nu < \nu_{\rm ssa} \right) 
&\sim
\frac{8 \pi^2}{3} \left( \frac{2 \pi m_e^3}{e} \right)^{1/2} \left(\frac{5}{6}\right)^{11/4} \left( 6 \sigma \right)^{-1/4} \nu^{5/2} \dot{E}^{3/10} \zeta^{-11/20} t^{61/20} 
\nonumber \\
&\simeq 8.50 \times 10^{28} \, {\rm erg \, s}^{-1} \, {\rm Hz}^{-1} \,
\nu_{10}^{2.5} \sigma_{-1}^{-0.25} \dot{E}_{41}^{0.3} \zeta_{-5}^{-0.55} t_{7}^{3.05}.
\end{align}
Here $\nu_{10} = \nu / 10^{10} \, {\rm Hz}$, $\sigma_{-1} = \sigma / 0.1$, $\dot{E}_{41} = \dot{E} / 10^{41} \, {\rm erg \, s}^{-1}$, and $t_7 = t / 10^7 \, {\rm s}$ denote quantities in cgs units, while $\zeta_{-5} = \zeta / \left[ 10^{-5} \left(M_{\rm ej}/M_\odot \right) \left( v_{\rm ej} / 10^4 \, {\rm km \, s}^{-1} \right)^{-3} \right]$.

The optically thin synchrotron emission depends directly on the population of emitting electrons as 
\begin{equation}
L_\nu \approx \frac{3 e^3}{m_e c^2} \gamma N_\gamma B_{\rm n} .
\end{equation}

Although no simple analytic solution for the electron distribution $N_\gamma$ exists in the case where both synchrotron and adiabatic cooling are important, we find in practice and in comparison with numerical calculations that
Equation~(\ref{eq:N_gamma_syn}) in the
regime where $\gamma \lesssim \gamma_{\rm cool}(t)$ reasonably approximates the electron distribution at times of interest (near peak light-curve in the GHz band) for a wide range of parameters, even when adiabatic cooling is included in addition to synchrotron cooling.

The optically-thin synchrotron flux in this regime is
\begin{align} \label{eq:Lnu_thin}
L_\nu \left( \nu > \nu_{\rm ssa} \right) &\sim 
\left(\frac{5}{8}\right)^{13/8} \left(\frac{6}{5}\right)^{93/32} \left( 32\pi \right)^{-5/16} \frac{3 e^{43/16} \sigma_{\rm T}^{5/8}}{m_e^{21/16} c^{37/16}} \left( 6\sigma \right)^{31/32}
\chi^{-1} \dot{E}^{111/80} \zeta^{93/160} t^{-143/160} \nu^{5/16}
\nonumber \\
&\simeq 4.46 \times 10^{28} \, {\rm erg \, s}^{-1} \, {\rm Hz}^{-1} \,
\nu_{10}^{0.31} \chi_{0.2{\rm GeV}}^{-1} \sigma_{-1}^{0.97} \dot{E}_{41}^{1.39} \zeta_{-5}^{0.58} t_{7}^{-0.89}
,
\end{align}
where $\chi_{0.2{\rm GeV}} = \chi / 0.2\,{\rm GeV}$.
Putting aside external free-free absorption by the ejecta for the moment, the nebula synchrotron luminosity is the minimum of Equations~(\ref{eq:Lnu_thick},\ref{eq:Lnu_thin}).
The intrinsic peak of the light-curve at a given frequency therefore coincides with the synchrotron self-absorption turnover, which occurs when the optically thick and optically thin synchrotron expressions are roughly equal. Equating (\ref{eq:Lnu_thick}) and (\ref{eq:Lnu_thin}), we find the time of the self-absorption turnover,
\begin{equation} \label{eq:t_ssa}
t_{\rm ssa} \simeq 8.49 \times 10^6 \, {\rm s} \, 
\nu_{10}^{-350/631} \sigma_{-1}^{195/631} \chi_{0.2{\rm GeV}}^{-160/631} \dot{E}_{41}^{174/631} \zeta_{-5}^{181/631}
.
\end{equation}
The corresponding peak luminosity can be obtained by plugging this timescale back into Equation~(\ref{eq:Lnu_thick}), giving
\begin{equation} \label{eq:Lnu_ssa}
L_{\nu}\left(t_{\rm ssa}\right) \simeq 5.16 \times 10^{28} \, {\rm erg \, s}^{-1} \, {\rm Hz}^{-1} \, 
\nu_{10}^{510/631} \sigma_{-1}^{437/631} \chi_{0.2{\rm GeV}}^{-488/631} \dot{E}_{41}^{720/631} \zeta_{-5}^{205/631}.
\end{equation}
Equations~(\ref{eq:t_ssa},\ref{eq:Lnu_ssa}) approximate the peak time and luminosity of the nebular radio emission if this peak occurs after the surrounding ejecta has already become transparent to free-free absorption.
Otherwise, the observed emission peaks at the time when the ejecta eventually does become free-free transparent at radio frequencies, $t_{\rm ff}$, and the nebular synchrotron emission is in the optically thin regime (\ref{eq:Lnu_thin}), i.e.
\begin{align} \label{eq:peak_radio}
t_{\rm pk} &= \max \left( t_{\rm ssa}, t_{\rm ff} \right)
\\
L_{\nu,{\rm pk}} &= 
\begin{cases}
L_{\nu}\left(t_{\rm ssa}\right) \,\,\, , \,\,\, t_{\rm pk} = t_{\rm ssa}
\\
L_{\nu>\nu_{\rm ssa}}\left(t_{\rm ff}\right) \approx L_{\nu}\left(t_{\rm ssa}\right) \left(\frac{t_{\rm ff}}{t_{\rm ssa}}\right)^{-143/160}
\,\,\, , \,\,\, t_{\rm pk} = t_{\rm ff}.
\end{cases}
\end{align}

Even for the simple model assumed above (constant energy injection rate, delta-function Lorentz factor injection of electrons) the expression derived in Equations~(\ref{eq:Lnu_thin},\ref{eq:t_ssa},\ref{eq:Lnu_ssa}) are only approximate estimates. This is due to approximations made in solving the electron Lorentz factor distribution function, which include among other things the neglect of adiabatic cooling, and inhibition of synchrotron cooling in the synchrotron self-absorption optically thick regime.
Nevertheless, we find that these estimates typically agree with our numerical simulations (which were used to construct models A and B illustrated in figures~\ref{fig:nebula},\ref{fig:radio}) to within a factor of $\sim$several in the range of parameters explored for our purposes.

\end{document}